# Detect adverse drug reactions for the drug Pravastatin

Yihui Liu[1,2]

[1]Institute of Intelligent Information Processing,Shandong
Polytechnic University, China
yihui_liu_2005@yahoo.co.uk

Uwe Aickelin[2]

[2]Department of Computer Science,
University of Nottingham, UK

*Abstract*—**Adverse drug reaction (ADR) is widely concerned for public health issue. ADRs are one of most common causes to withdraw some drugs from market. Prescription event monitoring (PEM) is an important approach to detect the adverse drug reactions. The main problem to deal with this method is how to automatically extract the medical events or side effects from high-throughput medical data, which are collected from day to day clinical practice. In this study we propose an original approach to detect the ADRs using feature matrix and feature selection. The experiments are carried out on the drug Pravastatin. Major side effects for the drug are detected. The detected ADRs are based on computerized method, further investigation is needed.**

## I. INTRODUCTION

Adverse drug reaction (ADR) is widely concerned for public health issue. ADRs are one of most common causes to withdraw some drugs from market [1]. Now two major methods for detecting ADRs are spontaneous reporting system (SRS) [2,3,4,5], and prescription event monitoring (PEM) [6,7,8,9,10]. The World Health Organization (WHO) defines a signal in pharmacovigilance as "any reported information on a possible causal relationship between an adverse event and a drug, the relationship being unknown or incompletely documented previously" [11]. For spontaneous reporting system, many machine learning methods are used to detect ADRs, such as Bayesian confidence propagation neural network (BCPNN) [12,13], decision support method [14], genetic algorithm [15], knowledge based approach [16,17], etc. One limitation is the reporting mechanism to submit ADR reports [14], which has serious underreporting and is not able to accurately quantify the corresponding risk. Another limitation is hard to detect ADRs with small number of occurrences of each drug-event association in the database. Prescription event monitoring has been developed by the Drug Safety Research Unit (DSRU) and is the first large-scale systematic post-marketing surveillance method to use event monitoring in the UK [8]. The health data are widely and routinely collected. It is a key step for PEM methods to automatically detect the ADRs from thousands of medical events. In paper [18,19], MUTARA and HUNT, which are based on a domain-driven knowledge representation Unexpected Temporal Association Rule, are proposed to signal unexpected and infrequent patterns characteristic of ADRs, using Queensland Hospital morbidity data, more commonly referred to as the Queensland Linked Data Set (QLDS) [20]. But their methods achieve low accuracies for detecting ADRs.

In this paper we propose feature selection approach to detect ADRs from The Health Improvement Network (THIN) database. First feature matrix, which represents the medical events for the patients before and after taking drugs, is created by linking patients' prescriptions and corresponding medical events together. Then significant features are selected based on feature selection methods, comparing the feature matrix before patients take drugs with one after patients take drugs. Finally the significant ADRs can be detected from thousands of medical events based on corresponding features. Experiments are carried out on the drug Pravastatin. Good performance is achieved.

## II. FEATURE MATRIX AND FEATURE SELECTION

### A. The THIN Database

The Health Improvement Network (THIN) is a collaboration product between two companies of EPIC and InPS. EPIC is a research organization, which provides the electronic database of patient care records from UK and other countries. There are 'Therapy' and 'Medical' databases in THIN data. The "Therapy" database contains the details of prescriptions issued to patients. Information of patients and the prescription date for the drug can be located. The "Medical" database contains a record of symptoms, diagnoses, and interventions recorded by the general practice (GP) and/or primary care team. Each symptom for patients forms a record. By linking patient identifier, their prescriptions, and their corresponding medical events (symptoms) together, feature matrix to characterize the symptoms during the period before or after patients take drugs is built.

### B. The Extraction of Feature Matrix

To detect the ADRs of drugs, first feature matrix is extracted from THIN database, which describes the medical events that patients occur before or after taking drugs. Then feature selection method of Student's t-test is performed to select the significant features from feature matrix containing thousands of medical events. Figure 1 shows the process to detect the ADRs using feature matrix. Feature matrix $A$ (Figure 2) describes the medical events for each patient during 60 days before they take drugs. Feature matrix $B$ reflects the medical events during 60 days after patients take drugs. In order to reduce the effect of the small events, and save the computation time and space, we set 100 patients as a group. Matrix $X$ and $Y$ are feature matrix after patients are divided into groups.

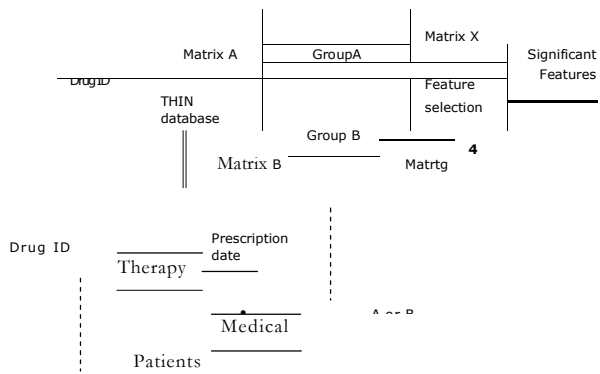

Figure 1. The process to detect ADRs. Matrix *A* and *B* are feature matrix before patients take drugs or after patients take drugs. The time period of observation is set to 60 days. Matrix *X* and Y are feature matrix after patients are divided into groups. We set 100 patients as one group.

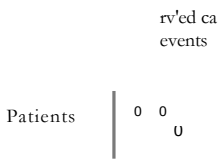

Figure 2. Feature matrix. The row of feature matrix represents the patients, and the column of feature matrix represents the medical events. The element of events is 1 or 0, which represents the patient having or not having the corresponding medical symptom or event.

## C. Medical Events and Readcodes

Medical events or symptoms are represented by medical codes or Readcodes. There are 103387 types of medical events in "Readcodes" database. The Read Codes used in general practice (GP), were invented and developed by Dr James Read in 1982. The NHS (National Health Service) has expanded the codes to cover all areas of clinical practice. The code is hierarchical from left to right or from level 1 to level 5. It means that it gives more detailed information from level 1 to level 5. Table 1 shows the medical symptoms based on Readcodes at level 3 and at level 5. 'Other soft tissue disorders' is general term using Readcodes at level 3. 'Foot pain', 'Heel pain', etc., give more details using Readcodes at level 5.

TABLE I.    TABLE I. MEDICAL EVENTS BASED ON READCODES AT LEVEL 3 AND LEVEL 5.

|  | Level | Readcodes | Medical events |
|---|---|---|---|
|  | Level 3 | N24..00 | Other soft tissue disorders |
| Muscle pain |  | N245.16 | Leg pain |
|  | Level 5 | N245111 | Toe pain |
|  |  | N245.13 | Foot pain |
|  |  | N245700 | Shoulder pain |
|  |  | N245.15 | Heel pain |

## D. Feature Selection Based on Student's t-test

Feature extraction and feature selection are widely used in biomedical data processing [21-27]. In our research we use Student's t-test [28] feature selection method to detect the significant ADRs from thousands of medical events. Student's t-test is a kind of statistical hypothesis test based on a normal distribution, and is used to measure the difference between two kinds of samples.

## E. Other parameters

The variable of ratio $R_1$ is defined to evaluate significant changes of the medical events, using ratio of the patient number after taking the drug to one before taking the drug. The variable $R_2$ represents the ratio of patient number after taking the drug to the number of whole population having one particular medical symptom.

The ratio variables $R_1$ and $R_2$ are defined as follows:

$$R_1 = [N_A / _{3TB} \text{ if } NB *0;$$

$$NA \quad if\ NB = 0;$$

$$R2 = NAI\ N$$

where $N_B$ and $N_A$ represent the numbers of patients before or after they take drugs for having one particular medical event respectively. The variable $N$ represents the number of whole population who take drugs.

The drug Pravastatin is used to test our proposed method, using 475 GPs data in THIN database. Student's t-test is performed to select the significant features from feature matrix, which represent the medical events having the significant changes after patients take the drugs.

Drugs.com provides access to healthcare information tailored for a professional audience, sourced solely from the most trusted, well-respected and independant agents such as the Food and Drug Administration (FDA), American Society of Health-System Pharmacists, Wolters Kluwer Health, Thomson Micromedex, Cerner Multum and Stedman's. In the Drugs.com [29], side effects for Pravastatin are listed, such as severe allergic reactions (rash; hives; itching; difficulty breathing; tightness in the chest; swelling of the mouth, face, lips, or tongue); change in the amount of urine produced; chest pain; dark urine; fever, chills, or persistent sore throat; flu-like symptoms; muscle pain, tenderness, or weakness (with or without fever or fatigue); pale stools; red, swollen, blistered, or peeling skin; severe stomach pain; vision changes; yellowing of the eyes or skin, etc.

There are seven currently prescribed forms of statin drugs. They are Rosuvastatin, Atorvastatin, Simvastatin, Pravastatin, Fluvastatin, Lovastatin, Pitavastatin, muscle pain and musculoskeletal events [29, 30] are two of the main side effects of statin drugs.

In experiments, we use two kinds of feature matrix. One feature matrix is based on all medical events using Readcodes at level 1-5 in order to observe the detailed symptoms. Another one is based on the medical events using Readcodes at level 13, combining the detailed information of Readcodes at level 4 and 5 into one general term using Readcodes at level 3. 10875 patients from half data of 475GPs are taking Pravastatin. 13438 medical events or symptoms are obtained using Readcodes at level 1-5, and the feature matrix of 10875x13438 is formed to describe the detailed information of patients. After grouping them, 108x13435 feature matrix is obtained. For Readcodes at level 1-3, 108x2764 feature matrix is obtained.

Table 2 shows the top 20 detected results in ascending order of p value of Student's t-test, using Readcodes at level 15 and at level 1-3. The detected results are using p values less than 0.05, which represent the significant change after patients take the drug. Table 3 shows the results in descending order of the ratio of the number of patients after taking the drug to one before taking the drug. Our detected results are consistent with the listed results in [29,30]. Table 4 shows cancer information related to the patients who take the drug Pravastatin.

IV. CONCLUSIONS

In this study we propose a novel method to successfully detect the ADRs using feature matrix and feature selection. A feature matrix, which characterizes the medical events before patients take drugs or after patients take drugs, is created from THIN database. The feature selection method of Student's t-test is used to detect the significant features from thousands of medical events. The significant ADRs, which are corresponding to significant features, are detected. The detected ADRs are based on our proposed method, further investigation is needed.

TABLE II.    THE TOP POTENTIAL ADRS FOR PRAVASTATIN BASED ON P VALUE OF STUDENT'S T-TEST.

| | Rank | Readcodes | Medical events | NB | NA | R1 | R2 |
|---|---|---|---|---|---|---|---|
| Level 1-5 | 1 | 1Z12.00 | Chronic kidney disease stage 3 | 84 | 532 | 6.33 | 5.14 |
| | 2 | C34..00 | Gout | 85 | 280 | 3.29 | 2.71 |
| | 3 | 1M10.00 | Knee pain | 102 | 421 | 4.13 | 4.07 |
| | 4 | 1969.00 | Abdominal pain | 81 | 389 | 4.80 | 3.76 |
| | 5 | A53..11 | Shingles | 16 | 153 | 9.56 | 1.48 |
| | 6 | 1D14.00 | C/O: a rash | 135 | 497 | 3.68 | 4.80 |
| | 7 | M03z000 | Cellulitis NOS | 37 | 252 | 6.81 | 2.43 |
| | 8 | N245.17 | Shoulder pain | 130 | 488 | 3.75 | 4.71 |
| | 9 | F4C0.00 | Acute conjunctivitis | 69 | 288 | 4.17 | 2.78 |
| | 10 | N131.00 | Cervicalgia - pain in neck | 129 | 419 | 325 | 4.05 |
| | 11 | 1C9..00 | Sore throat symptom | 61 | 280 | 4.59 | 2.71 |
| | 12 | C10F.00 | Type 2 diabetes mellitus | 183 | 507 | 2.77 | 4.90 |
| | 13 | F46..00 | Cataract | 33 | 219 | 6.64 | 2.12 |
| | 14 | H06z000 | Chest infection NOS | 132 | 688 | 5.21 | 6.65 |
| | 15 | N143.00 | Sciatica | 78 | 291 | 3.73 | 2.81 |
| | 16 | H01..00 | Acute sinusitis | 53 | 218 | 4.11 | 2.11 |
| | 17 | H060.00 | Acute bronchitis | 58 | 329 | 5.67 | 3.18 |
| | 18 | 16C6.00 | Back pain without radiation NOS | 107 | 407 | 3.80 | 3.93 |
| | 19 | J16v400 | Dyspepsia | 122 | 384 | 3.15 | 3.71 |
| | 20 | N245.13 | Foot pain | 36 | 205 | 5.69 | 1.98 |
| Level 1-3 | 1 | 171..00 | Cough | 473 | 1799 | 3.80 | 16.54 |
| | 2 | H06..00 | Acute bronchitis and bronchiolitis | 383 | 1574 | 4.11 | 14.47 |
| | 3 | N21..00 | Peripheral enthesopathies and allied syndromes | 184 | 685 | 3.72 | 6.30 |
| | 4 | 1Z1..00 | Chronic renal impairment | 139 | 758 | 5.45 | 6.97 |
| | 5 | M22..00 | Other dermatoses | 97 | 476 | 4.91 | 4.38 |
| | 6 | 1B8..00 | Eye symptoms | 125 | 542 | 4.34 | 4.98 |
| | 7 | N24..00 | Other soft tissue disorders | 750 | 1933 | 2.58 | 17.77 |
| | 8 | H05..00 | Other acute upper respiratory infections | 179 | 862 | 4.82 | 7.93 |
| | 9 | F4C..00 | Disorders of conjunctiva | 128 | 537 | 420 | 4.94 |
| | 10 | 19F..00 | Diarrhoea symptoms | 147 | 633 | 4.31 | 5.82 |
| | 11 | 1B1..00 | General nervous symptoms | 307 | 1052 | 3.43 | 9.67 |
| | 12 | N14..00 | Other and unspecified back disorders | 237 | 876 | 3.70 | 8.06 |
| | 13 | M03..00 | Other cellulitis and abscess | 79 | 404 | 5.11 | 3.71 |
| | 14 | 16C..00 | Backache symptom | 233 | 851 | 3.65 | 7.83 |
| | 15 | 1M1..00 | Pain in lower limb | 125 | 536 | 4.29 | 4.93 |
| | 16 | N09..00 | Other and unspecified joint disorders | 320 | 1058 | 3.31 | 9.73 |
| | 17 | K19..00 | Other urethral and urinary tract disorders | 179 | 725 | 4.05 | 6.67 |
| | 18 | 196..00 | Type of GIT pain | 149 | 588 | 3.95 | 5.41 |
| | 19 | 1D1..00 | C/O: a general symptom | 352 | 1126 | 320 | 10.35 |
| | 20 | ABO..00 | Dermatophytosis including tinea or ringworm | 87 | 393 | 4.52 | 3.61 |

Variable NB and NA represent the numbers of patients before or after they take drugs for having one particular medical event. Variable R1 represents the ratio of the numbers of patients after taking drugs to the numbers of patients before taking drugs. Variable R2 represents the ratio of the numbers of patients after taking drugs to the number of the whole population.

TABLE III.    THE POTENTIAL ADRs FOR PRAVASTATIN BASED ON DESCENDING ORDER OF R1 VALUE.

| | Rank | Readcodes | Medical events | NB | NA | R1 | R2 |
|---|---|---|---|---|---|---|---|
| Level 1-5 | 1 | F591.00 | Sensorineural hearing loss | 1 | 36 | 36.00 | 0.33 |
| | 2 | J574700 | Anal pain | 1 | 26 | 26.00 | 0.24 |
| | 3 | 16J4.00 | Swollen knee | 0 | 25 | 25.00 | 0.23 |
| | 4 | D010.00 | Pernicious anaemia | 1 | 25 | 25.00 | 0.23 |
| | 5 | M22z.11 | Actinic keratosis | 1 | 25 | 25.00 | 0.23 |
| | 6 | A080300 | Infectious gastroenteritis | 1 | 24 | 24.00 | 0.22 |
| | 7 | B49..00 | Malignant neoplasm of urinary bladder | 1 | 23 | 23.00 | 0.21 |
| | 8 | 1NO3.00 | C/O: dry skin | 1 | 22 | 22.00 | 0.20 |
| | 9 | F420600 | Non proliferative diabetic retinopathy | 0 | 21 | 21.00 | 0.19 |
| | 10 | F42y900 | Macular oedema | 1 | 20 | 20.00 | 0.18 |
| | 11 | K190311 | Recurrent UTI | 1 | 19 | 19.00 | 0.17 |
| | 12 | 173C.12 | SOBOE | 0 | 19 | 19.00 | 0.17 |
| | 13 | SKz..00 | Injury NOS | 1 | 18 | 18.00 | 0.17 |
| | 14 | 1CA2.11 | Voice hoarseness | 3 | 53 | 17.67 | 0.49 |
| | 15 | J578.00 | Colonic polyp | 2 | 35 | 17.50 | 0.32 |
| | 16 | SE3..11 | Arm bruise | 1 | 17 | 17.00 | 0.16 |
| | 17 | 16J7.00 | Swollen foot | 0 | 17 | 17.00 | 0.16 |
| | 18 | 1B16.11 | Agitated - symptom | 1 | 17 | 17.00 | 0.16 |

| | Rank | Readcodes | Medical events | NB | NA | R1 | R2 |
|---|---|---|---|---|---|---|---|
| | 19 | S64..13 | Head injury | 1 | 17 | 17.00 | 0.16 |
| | 20 | G86..11 | Lymphoedema | 1 | 16 | 16.00 | 0.15 |
| Level 1-3 | 1 | 161..00 | Appetite symptom | 1 | 34 | 34.00 | 0.31 |
| | 2 | SE3..00 | Contusion, upper limb | 1 | 28 | 28.00 | 0.26 |
| | 3 | B49..00 | Malignant neoplasm of urinary bladder | 1 | 27 | 27.00 | 0.25 |
| | 4 | F4B..00 | Corneal opacity and other disorders of cornea | 1 | 21 | 21.00 | 0.19 |
| | 5 | 1C7..00 | Snoring symptoms | 1 | 18 | 18.00 | 0.17 |
| | 6 | 1BD..00 | Harmful thoughts | 1 | 18 | 18.00 | 0.17 |
| | 7 | SKz..00 | Injury NOS | 1 | 18 | 18.00 | 0.17 |
| | 8 | D31..00 | Purpura and other haemorrhagic conditions | 2 | 33 | 16.50 | 0.30 |
| | 9 | M25..00 | Disorders of sweat glands | 2 | 33 | 16.50 | 0.30 |
| | 10 | HOz..00 | Acute respiratory infection NOS | 1 | 16 | 16.00 | 0.15 |
| | 11 | BB1..00 | [M]Epithelial neoplasms NOS | 1 | 15 | 15.00 | 0.14 |
| | 12 | F41..00 | Retinal detachments and defects | 1 | 15 | 15.00 | 0.14 |
| | 13 | G41..00 | Chronic pulmonary heart disease | 1 | 15 | 15.00 | 0.14 |
| | 14 | K08..00 | Impaired renal function disorder | 1 | 15 | 15.00 | 0.14 |
| | 15 | B7J..00 | Haemangiomas and lymphangiomas of any site | 1 | 15 | 15.00 | 0.14 |
| | 16 | 1NO..00 | Skin symptoms | 2 | 29 | 14.50 | 0.27 |
| | 17 | SA1..00 | Open wound of knee, leg and ankle | 0 | 14 | 14.00 | 0.13 |
| | 18 | S93..00 | Open wound of finger(s) or thumb | 0 | 14 | 14.00 | 0.13 |
| | 19 | S24..00 | Fracture of carpal bone | 1 | 14 | 14.00 | 0.13 |
| | 20 | SD7..00 | Superficial injury of foot and toe(s) | 1 | 14 | 14.00 | 0.13 |
| | 21 | 1A3..00 | Micturition stream | 3 | 39 | 13.00 | 0.36 |
| | 22 | B83..00 | Carcinoma in situ of breast and genitourinary system | 0 | 13 | 13.00 | 0.12 |
| | 23 | F17..00 | Autonomic nervous system disorders | 0 | 13 | 13.00 | 0.12 |
| | 24 | K11..00 | Hydronephrosis | 0 | 13 | 13.00 | 0.12 |
| | 25 | F33..00 | Nerve root and plexus disorders | 1 | 13 | 13.00 | 0.12 |
| | 26 | E26..00 | Physiological malfunction arising from mental factors | 1 | 13 | 13.00 | 0.12 |
| | 27 | A38..00 | Septicaemia | 2 | 24 | 12.00 | 0.22 |
| | 28 | K13..00 | Other kidney and ureter disorders | 0 | 12 | 12.00 | 0.11 |
| | 29 | F40..00 | Disorders of the globe | 0 | 11 | 11.00 | 0.10 |
| | 30 | F12..00 | Parkinson's disease | 3 | 32 | 10.67 | 0.29 |

TABLE IV.   THE POTENTIAL ADRs RELATED TO CANCER FOR PRAVASTATIN BASED ON P VALUE OF STUDENT S T-TEST.

| | Rank | Readcodes | Medical events | NB | NA | R1 | R2 |
|---|---|---|---|---|---|---|---|
| Cancer (level 1-5) | 1 | B33..11 | Basal cell carcinoma | 29 | 126 | 4.34 | 1.16 |
| | 2 | B46..00 | Malignant neoplasm of prostate | 8 | 47 | 5.88 | 0.43 |
| | 3 | B49..00 | Malignant neoplasm of urinary bladder | 1 | 23 | 23.00 | 0.21 |
| | 4 | BB2A.00 | [M]Squamous cell carcinoma NOS | 2 | 29 | 14.50 | 0.27 |
| | 5 | B34..00 | Malignant neoplasm of female breast | 7 | 30 | 4.29 | 0.28 |
| | 6 | BB2..12 | [M]Squamous cell neoplasms | 2 | 17 | 8.50 | 0.16 |
| | 7 | B32..00 | Malignant melanoma of skin | 0 | 9 | 9.00 | 0.08 |
| | 8 | B13..00 | Malignant neoplasm of colon | 2 | 15 | 7.50 | 0.14 |
| | 9 | B22z.00 | Malignant neoplasm of bronchus or lung NOS | 4 | 16 | 4.00 | 0.15 |
| | 10 | B10..00 | Malignant neoplasm of oesophagus | 1 | 9 | 9.00 | 0.08 |
| | 11 | BB5..11 | [M]Adenocarcinomas | 4 | 15 | 3.75 | 0.14 |
| | 12 | B22..00 | Malignant neoplasm of trachea, bronchus and lung | 0 | 6 | 6.00 | 0.06 |
| | 13 | B585.00 | Secondary malignant neoplasm of bone and bone marrow | 0 | 6 | 6.00 | 0.06 |
| | 14 | B141.00 | Malignant neoplasm of rectum | 0 | 6 | 6.00 | 0.06 |
| | 15 | B834.00 | Carcinoma in situ of prostate | 0 | 5 | 5.00 | 0.05 |
| | 16 | BB12.00 | [M]Carcinoma NOS | 0 | 5 | 5.00 | 0.05 |
| | 17 | BB31.00 | [M]Basal cell carcinoma NOS | 1 | 7 | 7.00 | 0.06 |
| | 18 | B49z.00 | Malignant neoplasm of urinary bladder NOS | 0 | 6 | 6.00 | 0.06 |
| | 19 | B133.00 | Malignant neoplasm of sigmoid colon | 0 | 4 | 4.00 | 0.04 |
| | 20 | BB5M.00 | [M]Tubular adenomas and adenocarcinomas | 0 | 4 | 4.00 | 0.04 |
| | 21 | B590.11 | Carcinomatosis | 0 | 4 | 4.00 | 0.04 |
| | 22 | BB13.00 | [M]Carcinoma, metastatic, NOS | 1 | 6 | 6.00 | 0.06 |
| | 23 | BBEF.11 | [M]Lentigo maligna | 0 | 3 | 3.00 | 0.03 |
| | 24 | B577.00 | Secondary malignant neoplasm of liver | 0 | 3 | 3.00 | 0.03 |
| | 25 | H51y700 | Malignant pleural effusion | 0 | 3 | 3.00 | 0.03 |
| | 26 | B222.00 | Malignant neoplasm of upper lobe, bronchus or lung | 0 | 3 | 3.00 | 0.03 |
| | 27 | BB5y000 | [M]Basal cell adenocarcinoma | 0 | 3 | 3.00 | 0.03 |
| | 28 | 1J0..00 | Suspected malignancy | 0 | 3 | 3.00 | 0.03 |
| | 29 | B134.00 | Malignant neoplasm of caecum | 0 | 3 | 3.00 | 0.03 |
| | 30 | BB2B.00 | [M]Squamous cell carcinoma, metastatic NOS | 0 | 3 | 3.00 | 0.03 |